\documentclass{book}
\def\shorttitle{DAMA results}
\usepackage{narosa}
\begin{document}
\title{Results on Dark Matter and $\beta\beta$ decay modes by DAMA at Gran Sasso} 

\author{R. Bernabei\dag\footnote{e-mail:rita.bernabei@roma2.infn.it}}

\affil{\dag Dip. di Fisica, Universit\`a di Roma ``Tor Vergata''
and INFN-Roma Tor Vergata, 00133 Roma, Italy}

\beginabstract

DAMA is an observatory for rare processes and 
it is operative deep underground at the Gran Sasso National Laboratory 
of the I.N.F.N. (LNGS). Here some arguments will be presented on 
the investigation on dark matter particles by annual modulation signature 
and on some of the realized double beta decay searches. 
\endabstract

\section{Introduction}

DAMA is an observatory for rare processes and 
it is operative deep underground at the Gran Sasso National Laboratory 
of the I.N.F.N. (LNGS). 
The main experimental set-ups are: i) DAMA/NaI ($\simeq$ 100 kg of highly
radiopure NaI(Tl)), which completed its data taking on July 2002 
\cite{Psd96,Mod1,Nim98,Mod2,Ext,Mod3,Sist,Sisd,Inel,Hep,RNC,ijmd,ijma,epj06,pep97,Bel99a,Simp99,cnc,D1,axion,qballs,supden,supclu};
ii) DAMA/LXe ($\simeq$ 6.5 kg liquid Kr-free Xenon enriched either in $^{129}$Xe or in $^{136}$Xe) 
\cite{Bel90,LXe-el1,LXe-DM1,LXe-DM2-1,Xe98,LXe-el2,LXe-el3,LXe-DM2-3,Ndn00,Perf,qxe,Bey03Xe,lxe_npa2,LXe-bb1,LXe-bb2};
iii) DAMA/R\&D, devoted to tests on prototypes and to small scale experiments, 
mainly on the investigations of double beta decay modes 
in various isotopes \cite{lacl3,cnc_lacl3,eudec,Ber97,ca40,C2,cd106,ca48,Ce,Ba};
iv) the new second generation DAMA/LIBRA set-up ($\simeq$ 250 kg
highly radiopure NaI(Tl)) in operation since March 2003 \cite{vulcano};
v) the low background DAMA/Ge detector mainly devoted to sample measurements;
in some measurements the low-background Germanium detectors of the LNGS facility 
are also used \cite{armonia,lieubo}.
Moreover, a third generation R\&D is in 
progress towards a possible 1 ton set-up, DAMA proposed in 1996.

\section{The investigation on Dark Matter particles by annual modulation signature}

\subsection{The DAMA/NaI model independent result}
The highly radio-pure DAMA/NaI set-up \cite{Nim98,Sist,RNC,ijmd} has been a pioneer Dark Matter experiment of suitable exposed mass, 
sensitivity and stability 
of the running conditions,
taking data at LNGS over seven annual cycles. The main aim of DAMA/NaI has been the investigation
of the presence of Dark Matter (DM) particles in the
galactic halo by exploiting the model independent annual modulation signature,
originally suggested by \cite{Fre1,Fre2} in the middle of 80's. The data taking has been completed in July 2002 and 
still results are produced by various kinds of studies. The final model independent result on Dark Matter particles by DAMA/NaI has 
been published in \cite{RNC,ijmd}, while some of the many possible corollary model dependent investigations have been published in 
\cite{RNC,ijmd,ijma,epj06}; 
other corollary quests are available in literature, such as e.g.
refs. \cite{Bo03,Bo04,Botdm,khlopov,Wei01,foot,Saib}. Moreover, many other scenarios can be considered as well and some of them are 
already under analysis.
In addition, profiting by its low-background features and by the high collected exposure,
several results have been achieved both on Dark Matter particle investigations
with other approaches and on several other rare processes 
\cite{Psd96,pep97,Bel99a,Simp99,cnc,D1,axion,qballs,supden,supclu}.

We just remind that the annual modulation signature is based on the yearly Earth motion around the Sun, which is moving in the 
galactic halo. This motion induces 
a "wind" of Dark Matter particles on the Earth, 
whose intensity -- measured by the experimental setup 
deep underground -- varies along the year. This intensity 
should satisfy the following peculiarities: i) it varies as a cosine function; ii) with 1 year period; iii) with maximum around 
roughly ≈2 June; iv) with amplitude  not exceeding 7\% in many of the possible scenarios; but some scenarios exist which can expect 
a larger value. Moreover, the variation should be: v) present only in the low energy region where Dark Matter particles can induce the 
signal; vi) present only in events where energy is deposited only in one of the several scintillators of the setup since the 
probability of multiple interactions is negligible. The highly radiopure NaI(Tl) is particularly suitable for this kind of experiment 
since it is sensitive to Dark Matter particles of various natures, both of low (through Na target) and high mass (through I target) 
and with various type of interactions; moreover, it also offers many technical-practical advantages. In the DAMA/NaI experiment 
advanced techniques for low background experiments have been exploited and further developed; a continuous monitoring and control of 
all the running conditions was also assured.

This signature, when analysing the measured counting rate in a suitable large mass low background NaI(Tl), offers the possibility to 
point out the presence of Dark Matter particles in the galactic halo in a model independent way.

\begin{figure}[ht]
\centerline{\includegraphics[height=4.cm]{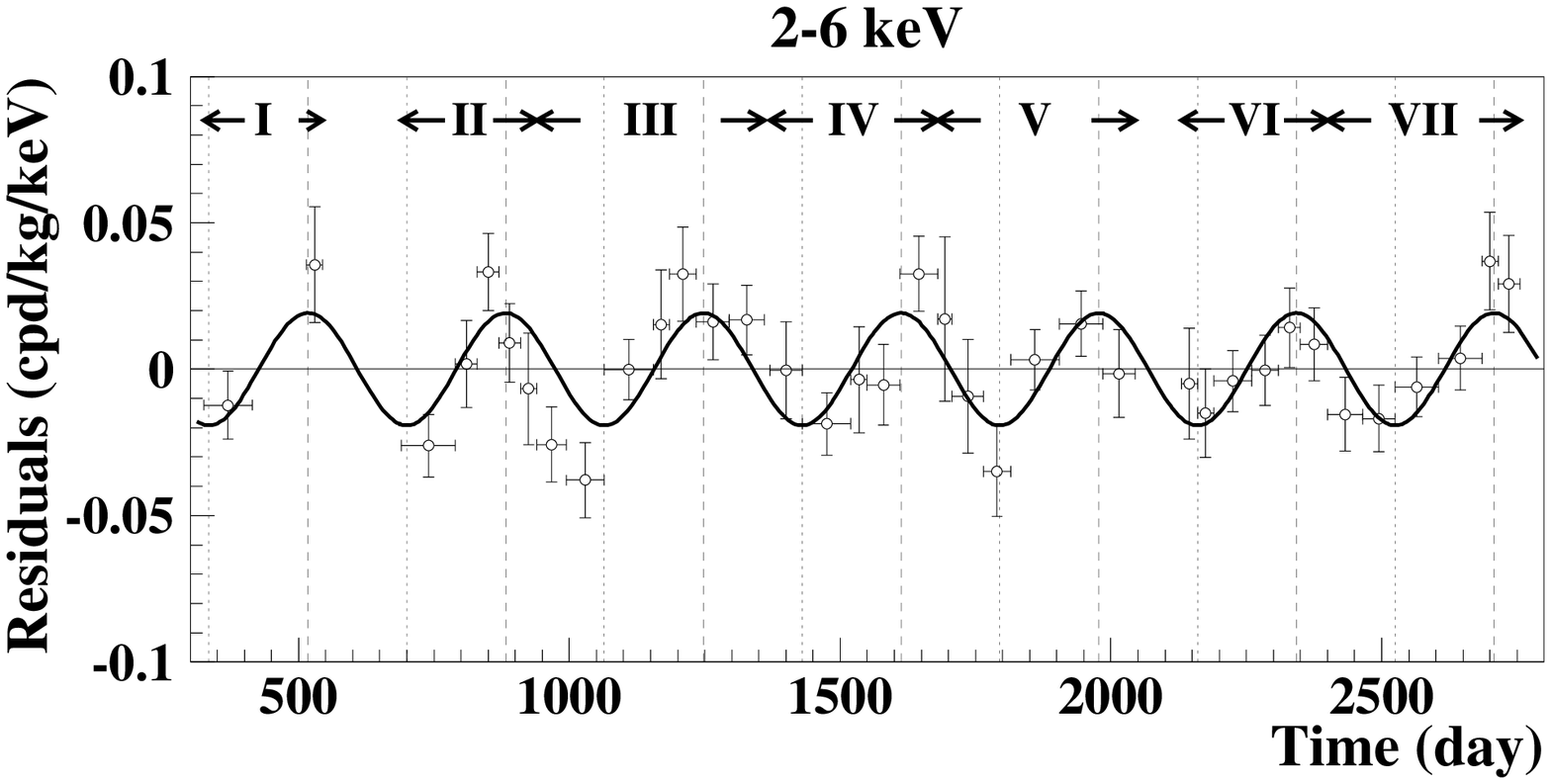}
            \includegraphics[height=4.cm]{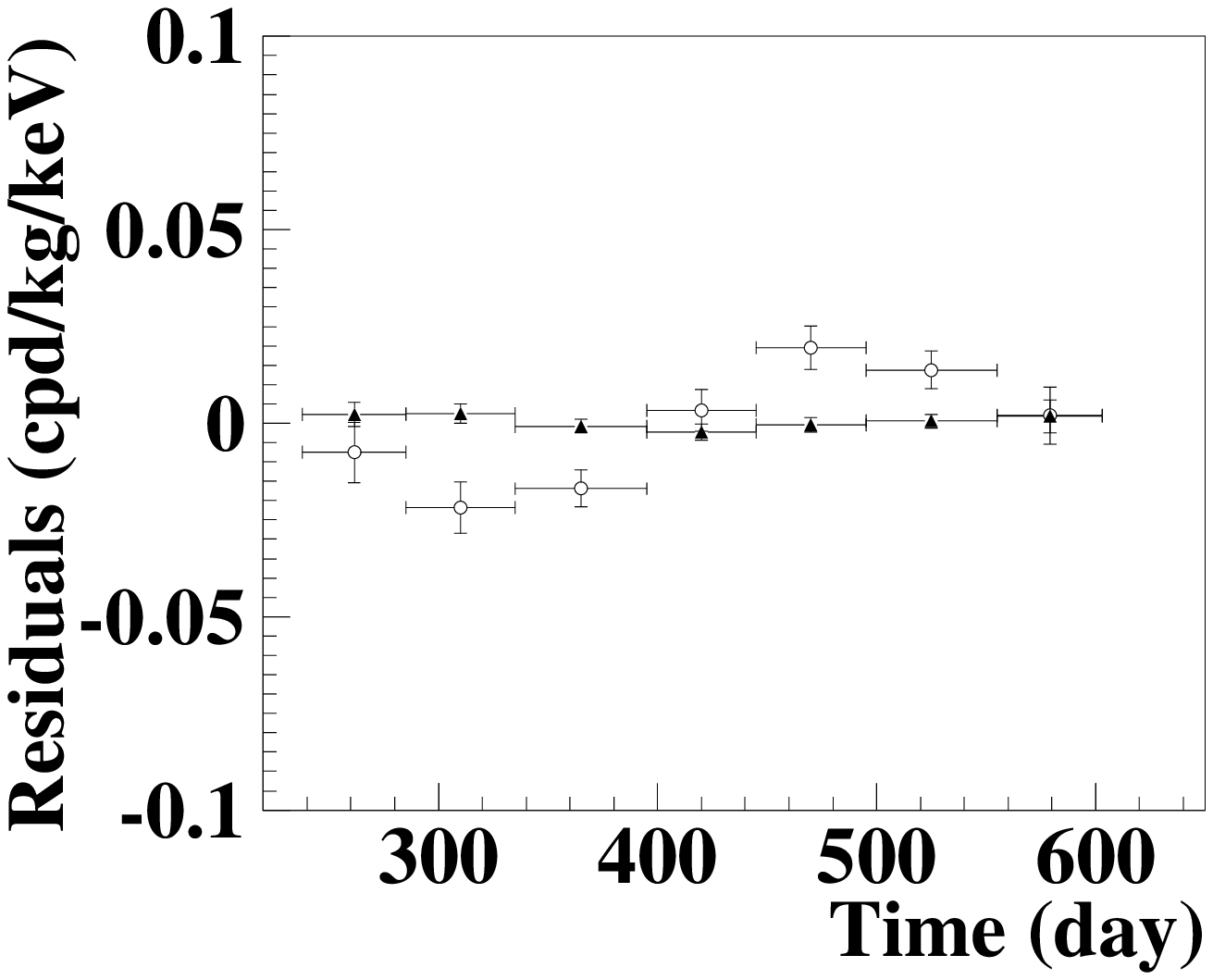}}   
\caption{{\em Left:} experimental
residual rate for {\em single-hit} events in the cumulative 
(2--6) keV energy interval
as a function of the time over 7 annual cycles, end of data taking July 2002.
The experimental points present the errors as vertical
bars and the associated time bin width as horizontal bars. The
superimposed curve represents the cosinusoidal function
behaviour expected for a Dark Matter particle signal with a period equal 
to 1 year and phase exactly at $2^{nd}$ June; the modulation
amplitude has been obtained by best fit \cite{RNC,ijmd}.
{\em Right:} experimental residual rates over seven annual cycles
for {\it single-hit} events (open circles) -- class of events
to which DM events belong --
and over the last two annual cycles for {\it multiple-hits}
events (filled triangles)
-- class of events to which DM events do not belong -- in the
(2--6) keV cumulative energy interval. They have been obtained
by considering for each class of events the data as collected
in a single annual cycle and using in both cases the same
identical hardware and the same identical software
procedures. The initial time is taken on August 7$^{th}$. 
This latter result offers an additional strong support for the presence of 
DM particles in the galactic halo further excluding any side effect 
either from hardware, from software procedures or from background.}
\label{fig1}
\end{figure}

In particular, the DAMA/NaI set-up has pointed out the presence of an annual modulation in the
{\it single-hit} residual rate in the lowest energy interval (2 -- 6) keV
(see Fig. \ref{fig1}); the observed effect satisfies the many peculiarities of the 
DM particle induced signature. This
gives a 6.3 $\sigma$ C.L. evidence over seven
annual cycles (total exposure: 107731 kg $\times$ day)
for the presence of DM particles in our Galaxy 
(see refs. \cite{Mod1,Mod2,Ext,Mod3,Sist,Sisd,Inel,Hep,RNC,ijmd,ijma,epj06};
moreover, further works on the corollary model dependent investigations are in progress). 
Neither systematic effects nor side reactions able to account for the 
measured modulation amplitude and to contemporaneously satisfy all the peculiarities 
of the signature have been found or suggested by anyone.

This result is model-independent. It represents the first 
experimental evidence of the presence of DM
particles in the galactic halo independently on their composition and nature.
At present, apart from DAMA/LIBRA, no other experiment is sensitive, 
for mass, radiopurity and control of the stability, to such a model independent signature
and, in particular, no other activity is available whose result can directly be compared 
in a model independent way with the one of DAMA/NaI.

\subsection{Some of the possible corollary quests for the DM candidate particles}

Several corollary investigations have been
pursued on the nature of the DM candidate particles
and on the astrophysical scenarios.
These corollary investigations are instead model-dependent
and, due to the poor knowledge on the 
astrophysical, nuclear and particle physics
needed assumptions and on the related parameters, they have no general
meaning  as it is also the case of exclusion plots and of DM 
particle parameters evaluated in indirect detection experiments. 
Thus, these investigations should be handled in the most general way (see for discussions at some extent e.g. 
\cite{RNC,ijmd,ijma,epj06} and in literature; other works are in progress). 

Several (of the many possible) corollary quests for a candidate particle have 
been carried out on the class of DM candidate particles named WIMPs \cite{RNC,ijmd}.
Low and high WIMP mass candidates interacting with ordinary matter via the general case of  mixed SI\&SD coupling and the subcases of 
dominant SI coupling or of dominant SD coupling
has been considered as well as the case of WIMPs with preferred SI inelastic 
scattering \cite{RNC,ijmd}.
In fact, the general solution for WIMPs is a 4-dimensional allowed volume in the space 
($\xi\sigma_{SI}$,$\xi\sigma_{SD}$,$\theta$,$m_W$) in the first case (being the two subcases
given by slices of this volume) and a 3-dimensional allowed volume in the space 
($\xi\sigma_{p}$,$\delta$,$m_W$) for the last one \cite{RNC,ijmd}. There, $\xi$ is the fractional amount of local density 
of DM 
particles, $\sigma_{SI}$ and $\sigma_{SD}$ are the point-like SI and SD DM 
particle-nucleon cross
sections and tg$\theta$ is the ratio between the effective SD coupling strengths of the
DM particle with neutron and proton ($\theta$ is defined in the [0,$\pi$) interval), $\sigma_{p}$
is the cross section for 
WIMPs with preferred inelastic interactions and $\delta$ is the mass splitting of the two mass states of this candidate.
 
\begin{figure}[ht]
\vspace{-.3cm}
\centerline{
\includegraphics[height=7.cm]{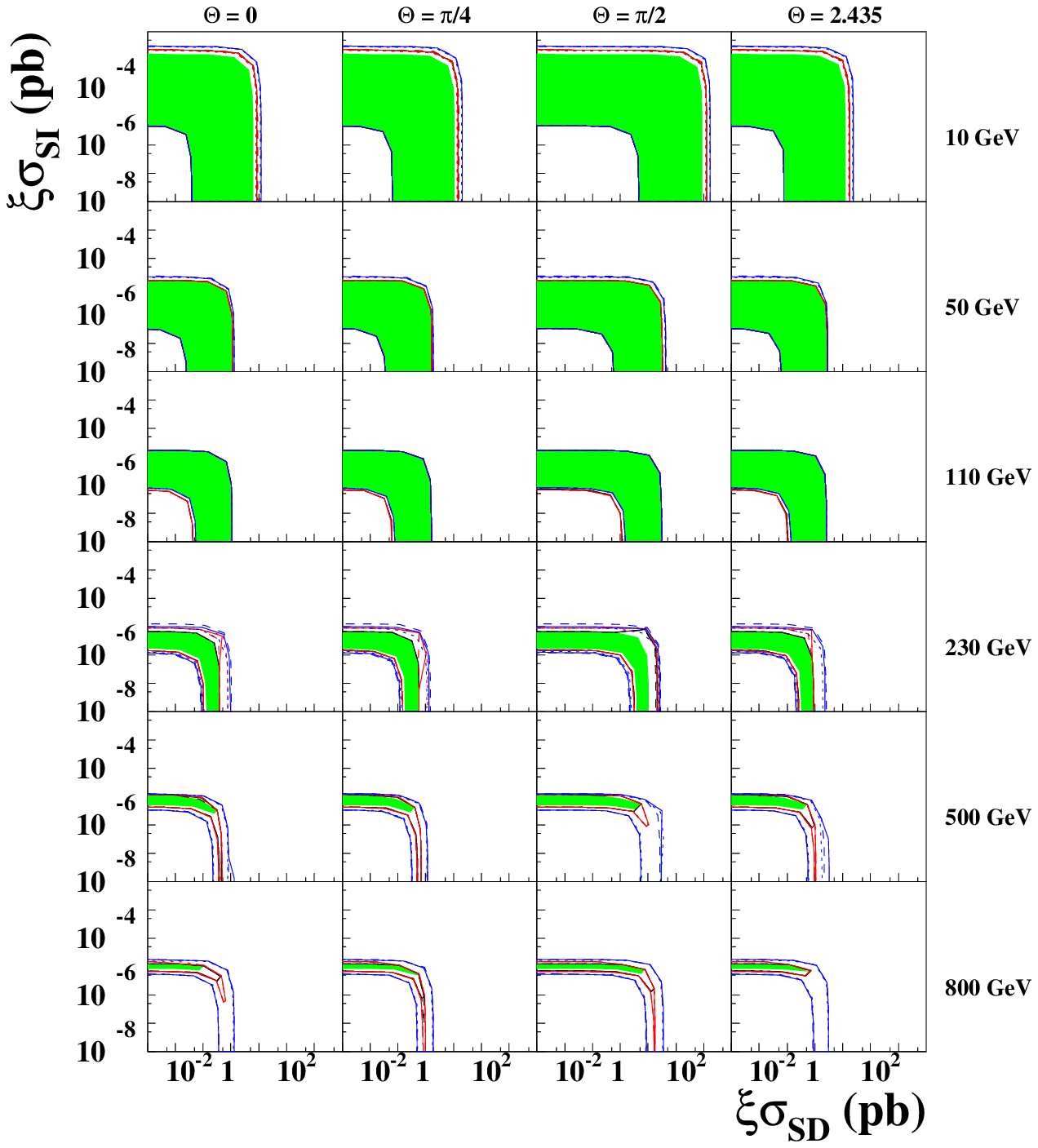}
\includegraphics[height=6.cm]{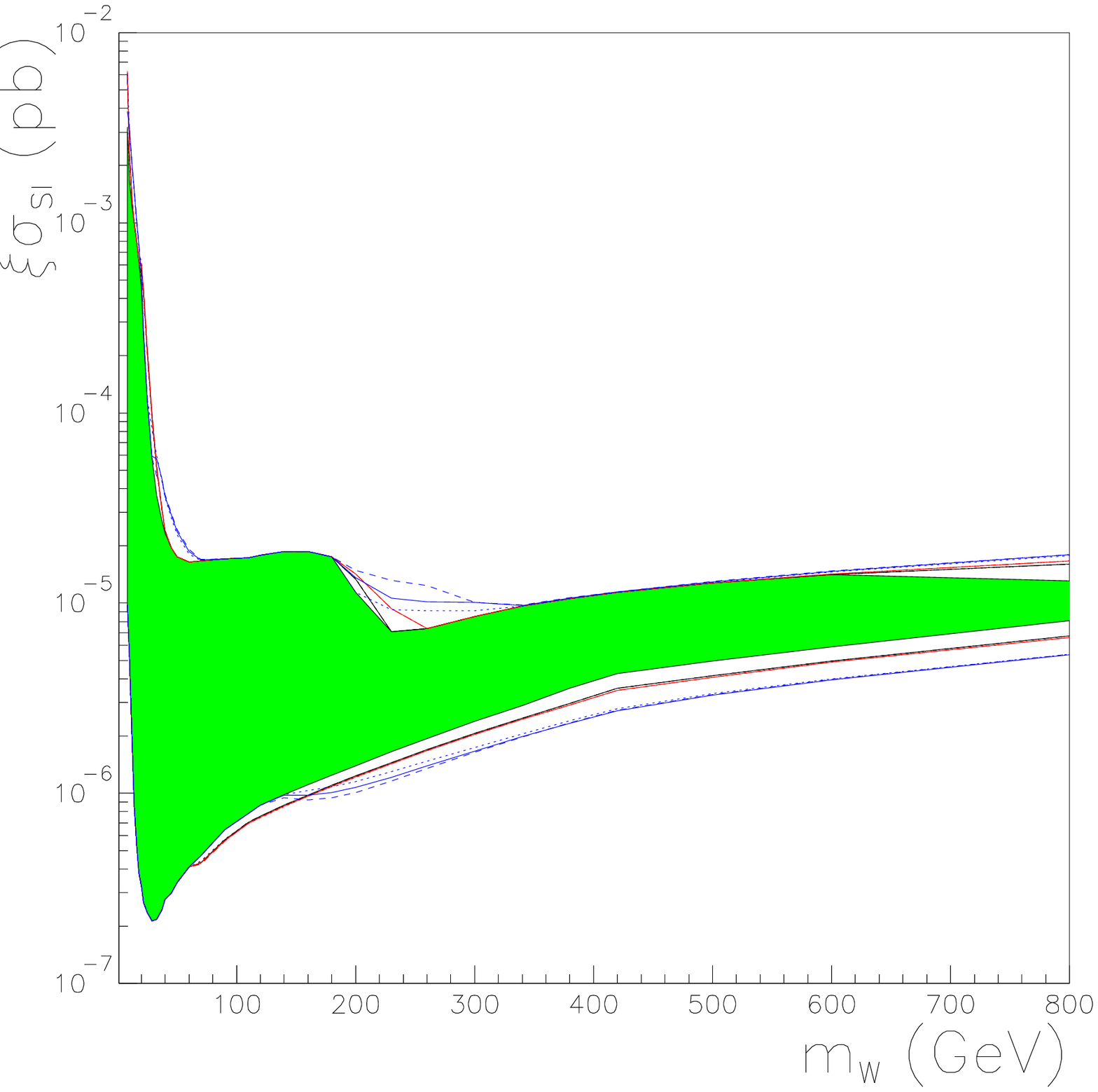}}
\vspace{-.2cm}
\caption{Left: filled areas are examples of slices, in the plane  $\xi \sigma_{SI}$ vs
$\xi \sigma_{SD}$, of the
4-dimensional allowed volume for DM particle with mixed SI\&SD interaction for the 
model frameworks given in \cite{RNC,ijmd}.
Right: filled area is the region allowed in the ($\xi\sigma_{SI},m_W$)
plane in the scenarios for pure SI coupling considered in \cite{RNC,ijmd}.
The areas enclosed by the lines are obtained by introducing the SagDEG contribution
\cite{epj06}.
In all the cases inclusion of other existing uncertainties on parameters and models
would further extend the regions and increase the sets of best fit values.}
\label{fg:fig3}
\vspace{-0.2cm}
\end{figure}

Accounting at some extent for the present poor knowledge of all the astrophysical, nuclear and particle Physics aspects the results 
have been given in terms of external contour obtained by the superposition of the allowed volumes/regions, that is they account for a 
large number of sets of best fit values. In particular, 
for the case of WIMP elastically scattering on nuclei
the obtained best-fit mass ranges from GeV to 
about 1 TeV and the cross section spans over more than one order of magnitude when accounting at least at some extent for some of the 
present uncertainties on assumptions and parameters. 
For some details and results see refs. \cite{RNC,ijmd}.
Furthermore, Galaxy hierarchical formation theories \cite{Navarro96}, 
numerical simulations \cite{moore}, the                                    
discovery of the Sagittarius Dwarf Elliptical Galaxy (SagDEG) in 1994 \cite{Ibata}
and more recent investigations \cite{bellaz}
suggest that the dark halo of the Milky Way                   
can have a rich phenomenology containing non thermalized
dark matter fluxes. 
Thus, recently the contribution of the
SagDEG has been investigated by analyzing the consequences of its DM stream 
contribution to the galactic halo on the DAMA/NaI annual modulation 
data.
For simplicity, in this analysis we have investigated these topics just considering
some of the candidates belonging to the WIMP class and few of the many possible 
astrophysical, nuclear and particle Physics scenarios as already considered 
in refs. \cite{RNC,ijmd} in order to allow a direct comparison. Some examples are given in 
Fig. \ref{fg:fig3}, while results on 
further aspects and details can be found in ref. \cite{epj06}. 

Other streams can potentially play 
more intriguing roles in the corollary investigations for whatever candidate particle
with whatever approach and for comparisons; they will be investigated in near future, such as 
e.g. the Canis Major \cite{bellaz} and
those arising from caustic halo models \cite{Sikivie}. 

The DAMA/NaI allowed volumes/regions are well compatible with
theoretical expectations for neutralino in various scenarios (see as an example \cite{Botdm}).
It is worth to remind that the neutralino itself 
has both SI and SD coupling and that the basic theory has 
a very large number of "open" parameters.
Also several other WIMP candidates fit the DAMA/NaI model independent evidence as well as
other existing candidates with a phenomenology similar, 
but not identical as for the WIMP cases e.g.: 
the mirror Dark Matter particles \cite{foot},
the self-interacting dark matter particles \cite{Saib}, etc. etc. Moreover, several candidates
not belonging to the WIMP class and having different interaction features, to which DAMA/NaI -- on the contrary of others -- 
is fully sensitive, are also available.
In addition, in principle whatever 
particle with suitable characteristics, even not yet foreseen 
by theories, can be a good candidate as 
DM in the galactic halo.

As a relevant case, the DAMA/NaI annual modulation data have also been analysed in terms 
of a light ($\simeq$ keV mass) bosonic candidate, either with
pseudoscalar or with scalar axion-like coupling; details are available in ref. \cite{ijma}.  
\begin{figure}[ht]
\centering
\vspace{-0.4cm}
\includegraphics[width=290pt] {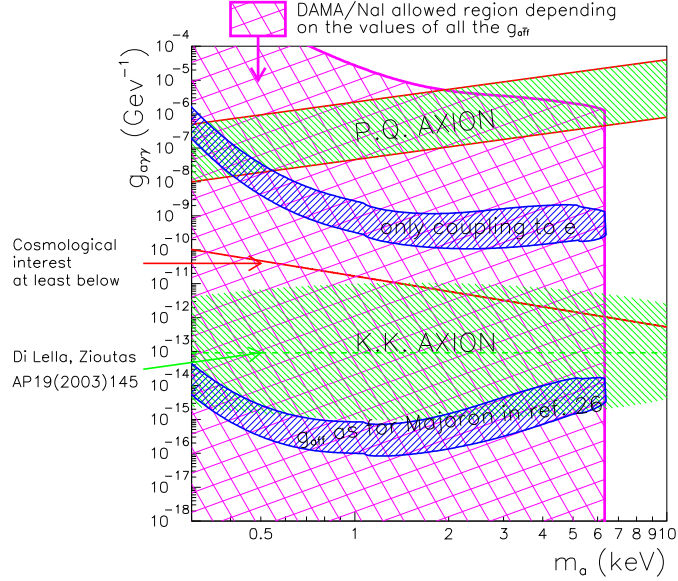}
\vspace{-0.4cm}
\caption{DAMA/NaI allowed region at 3$\sigma$ C.L. in the plane $g_{a\gamma\gamma}$ vs $m_a$
for a light pseudoscalar candidate ({\em crossed hatched region}).
All the configurations in this region can be allowed
depending on the values of all the $g_{a \bar{f}f}$.
Examples of two of the many possible models: i) {\em upper black region}: coupling only to electrons;
ii) {\em lower black region}: coupling (proportional to the $m_a$)
through the (weak) isospin to 
quarks and leptons. For details see ref. \cite{ijma}.
The indicative region of the Kaluza-Klein (K.K.) pseudoscalar axion credited in ref. \cite{DiLella} from the analysis
of indirect observations and the region of the DFSZ and KSVZ models (P.Q. axion) are shown as well.
The solid line corresponds to $a$ particle with lifetime equal to the
age of the Universe; at least all the $g_{a \gamma\gamma}$'s below this line are of cosmological interest.
However, in principle, it might be possible that the configurations above this line would also become
of interest in case of some exotic mechanism of the $a$ particle production.
Thus, a pseudoscalar DM candidate can also account for the DAMA/NaI model independent result as
well as the WIMP solutions already discussed elsewhere.}
\label{boso}
\end{figure}
For these candidates, the direct detection process is based on the total
conversion in NaI(Tl) crystal of the mass of the absorbed bosonic particle into
electromagnetic radiation. Thus, in these processes the target nuclei recoil is
negligible and is not involved in the detection process. Thus, signals from
these light bosonic DM candidates are detected in DAMA/NaI, but are lost in activities based 
on the selection/rejection procedures of 
the electromagnetic contribution to their counting rate
(such as e.g. Cdms, Edelweiss, Cresst, Warp, etc.) \cite{ijma}. Despite these particles are unstable,
their lifetime can be of cosmological interest and offers valuable candidates 
for the DM signal, observed in DAMA/NaI. In particular, the result for the pseudoscalar case 
for the same astrophysical scenarios of ref. \cite{RNC,ijmd} is shown in Fig. \ref{boso}.

Obviously, the discussed corollary quests are not at all exhaustive of the many scenarios
open at present level of knowledge for these and for other classes of candidates
as well as for the astrophysical and nuclear/atomic aspects.
Several other investigations have been discussed in literature and other studies 
are in progress.

\subsection{Few words on comparisons in the field}

No experiment is available -- with the exception of the presently running DAMA/LIBRA -- 
whose result can 
be directly compared in a model independent way with the experimental positive 6.3 $\sigma$ C.L. evidence 
for the presence of particle Dark Matter in the galactic halo obtained by DAMA/NaI.
Thus, claims for contradictions have intrinsically no scientific meaning.

In particular, as regards some claimed model-dependent comparisons presented so far we just mention 
-- among the many existing arguments -- that: 
i)     the other experiments are insensitive to the annual modulation signature; 
ii)    they release just a marginal exposure (orders of magnitude lower than the one by 
       DAMA/NaI) after several/many years underground; 
iii)   they exploit strong data selection and strong and often unsafe rejection techniques 
       of their huge counting rate, becoming at the same time insensitive to several 
       DM candidates; 
iv)    they generally quote in an incorrect/partial/not updated way the DAMA/NaI result;
v)     they consider a single model fixing all the astrophysical, nuclear and particle 
       Physics assumptions as well as all the theoretical and experimental parameters at a single 
       questionable choice \footnote{We note that the naive and partial 
       ``prescription'' of ref. \cite{smi96} on some aspects for a single particular WIMP case cannot be defined 
-- on the contrary of what appears in some paper -- as a ``standard theoretical model''. 
Such a paper summarized a single oversimplified approach adopted at that time.
Its use as ``unique'' reference is obviously incorrect, since it did
not account at all for the level of knowledge on all the involved astrophysics,
nuclear and particle physics aspects and parameters, for the many possibilities 
open on the astrophysical, nuclear and particle physics aspects and for the different existing approaches as e.g. the annual 
modulation signature. It is also worth to note that -- in addition -- the figure they usually reported for comparison is not only for 
the reasons, discussed here and elsewhere, 
in the substance meaningless, but it is intrisencally wrong also because e.g.: while they do not 
consider the model dependent LEP limit in their exclusion plots, they do not remove it from their "choice" of DAMA/NaI allowed space. 
If done, this will show -- as known since time (see literature) -- that even under their single arbitrary 
and naive scenario (which in addition cannot be applied at all to the annual modulation data which requires
-- among others -- time/energy correlation analysis), they are 
not discriminative at all.}.
Thus, e.g. for the WIMP case they do not account for the existing 
uncertainties on the real coupling with ordinary matter, on the
spin-dependent and spin-independent form factors and related parameters for each nucleus, 
on the spin factor used for each nucleus, on the real scaling laws for nuclear cross sections 
among different target materials; on the experimental and theoretical parameters, on
the effect of different halo models and related parameters on the different target materials, etc.
For example, large differences are expected in 
the counting rate among nuclei 
fully sensitive to the SD interaction (as $^{23}$Na and $^{127}$I)
with the respect to nuclei largely insensitive to such a coupling 
(as e.g $^{nat}$Ge, $^{nat}$Si, $^{nat}$Ar, $^{nat}$Ca, 
$^{nat}$W, $^{nat}$O) and also when nuclei in principle all sensitive 
to this coupling but having different 
unpaired nucleon (e.g. neutron in case of the odd spin nuclei, 
such as $^{129}$Xe, $^{131}$Xe, $^{125}$Te, $^{73}$Ge, $^{29}$Si, $^{183}$W 
and proton in the $^{23}$Na and $^{127}$I). Moreover, in case the 
detection of the DM particles would involve electromagnetic signals (see, for example,
the case of the light bosons discussed above, but also electromagnetic
contribution in WIMP detection arising e.g. from known effect induced by recoiling nuclei),
all the experiments, such as e.g. Cdms, Edelweiss, Cresst, Zeplin, Warp and 
their extensions, do lose the signal in their 
rejection procedures of the e.m. contribution to the counting rate.
In addition, those experiments present many critical points e.g. both regarding the energy scale
and the multiple selection procedures (related stabilities and efficiencies, systematics in the claimed
rejection factors, ranging from $10^{-4}$ to $10^{-8}$, evaluations, stabilities and monitoring of the
spill-out factors, ...) on which their claimed sensitivities (in a ``single'' set of assumptions 
and parameters' values) are based.

For completeness, it is also worth to note that no results obtained with different target material
can intrinsically
be directly compared even for the same kind of coupling, although
apparently all the presentations generally refer to cross section on the
nucleon. In fact, this requires -- among others -- 
the knowledge of e.g. the real scaling laws, form factors, etc.; the situation is much worse than that in the field 
of double beta decay experiments when different isotopes are used.

Discussions at some extent are also reported e.g. in ref. \cite{RNC,ijmd,ijma,epj06}; 
all those general comments also hold in the 
substance for more recent data releases.

As regards the indirect searches, a comparison 
would always require the calculation and the consideration of 
all the possible DM particle configurations in the
given particle model, since it does not exist
a biunivocal correspondence between the observables in the
two kinds of experiments. However, the present positive hints provided by indirect searches
are not in conflict with the DAMA/NaI result.

Finally, it is worth to note that -- among the many corollary aspects still open --
there is f.i. the possibility that the particle dark halo can have more than one component.
The more sensitive NaI(Tl) experiments in progress and in preparation by DAMA 
will also have this aspect among the arguments to be further investigated.

\subsection{The new DAMA/LIBRA and beyond}

In 1996 DAMA proposed to realize a ton set-up and a new R\&D project 
for highly radiopure NaI(Tl) detectors was funded.
As a consequence of the results of this second generation R\&D, 
the new experimental set-up DAMA/LIBRA (Large sodium Iodide Bulk for RAre processes),
$\simeq$250 kg highly radiopure NaI(Tl) crystal scintillators  
(matrix of twenty-five $\simeq$ 9.70 kg NaI(Tl) crystals), was funded and realised.
This set-up has replaced the previous DAMA/NaI; the experimental site 
as well as many components of the installation itself 
have been implemented (environment, shield of the photomultipliers, 
wiring, High-Purity Nitrogen system, 
cooling water of air conditioner, electronics and data acquisition system, etc.). 
In particular, all 
the Copper parts have been chemically etched before their installation 
following a new devoted protocol and all the procedures performed during the 
dismounting of DAMA/NaI 
and the installation of DAMA/LIBRA detectors have been carried out in 
High-Purity Nitrogen atmosphere \cite{vulcano}.

DAMA/LIBRA is continuously taking data
since March 2003; for example, in March 2007,
an exposure of about $1.5 \times 10^5$ kg $\times$ day has already been collected.
As in the case of the DAMA/NaI experiment regular routine 
calibrations are performed, e.g. in 4 years of measurements about
$ 4 \times 10^7$ events of sources for energy calibration and
$ 2 \times 10^6$ events/keV for the evaluation of the acceptance window efficiency 
have been acquired. 

A large work is going to be faced by the new DAMA/LIBRA, which will also further investigate 
with higher sensitivity several other rare processes.
The first data release is foreseen not later than 2008.

A third generation R\&D effort towards a possible NaI(Tl) ton set-up, we proposed in 1996, 
has been funded by I.N.F.N. and related works on 
selection of materials for detectors and photomultipliers are in progress \cite{dama1ton}.

\section{The searches for double beta decay modes}

\begin{figure}[ht]
\centering 
\vspace{-0.4cm}
\includegraphics[width=0.65\textwidth,angle=-90]{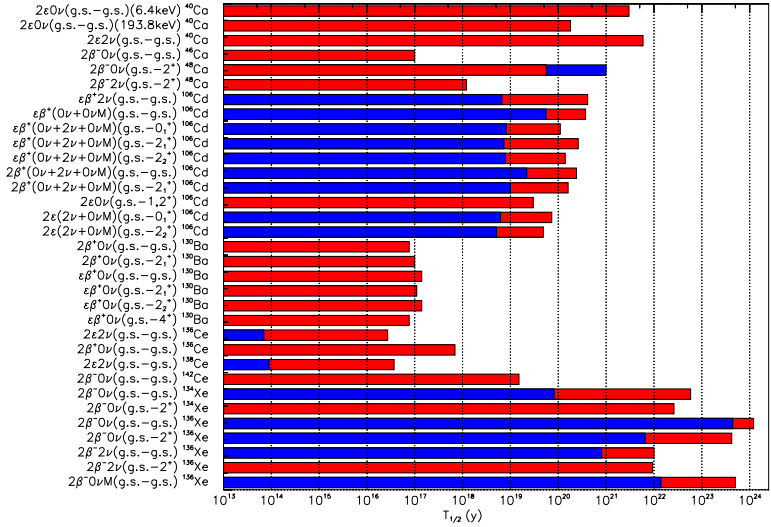}
\vspace{-0.2cm}
\caption{Summary of the $T_{1/2}$ limits 
 obtained by DAMA (red bars) and by previous experiments 
(blue bars) on  various double beta decay processes
All limits are at 90\% C.L. except  the 
$2\beta^+0\nu$ of $^{136}$Ce and $2\beta^+0\nu$ of $^{142}$Ce 
that are at 68\% C.L..}
\label{sum}
\end{figure}

The various low background set-ups, realized by the DAMA collaboration, also allow 
to investigate several kinds of rare processes.
In particular, several double beta decay modes have been investigated 
by using either the active or the passive source technique 
or -- sometimes -- the coincidence technique.
Fig. \ref{sum} summarizes the many results obtained by DAMA in 
these searches \cite{LXe-bb1,LXe-bb2,Ber97,ca40,C2,cd106,ca48,Ce,Ba}.

In particular, for this purpose the DAMA/LXe set-up 
\cite{LXe-bb1,LXe-bb2} and the DAMA/R\&D set-up \cite{Ber97,ca40,C2,cd106,ca48,Ce,Ba} 
have been used, as well as the low-background Ge in the LNGS facility \cite{armonia,lieubo}.
Among the obtained results,  we remind the search for:
i) $\beta\beta$ decay modes in $^{136}$Xe and in $^{134}$Xe \cite{LXe-bb1,LXe-bb2};
ii) $\beta\beta$ decay modes in $^{136}$Ce and in $^{142}$Ce \cite{Ber97}; 
iii) 2EC2$\nu$ decay mode in $^{40}$Ca \cite{ca40};
iv) $\beta\beta$ decay modes in $^{46}$Ca and in $^{40}$Ca \cite{C2};
v) $\beta\beta$ decay modes in $^{106}$Cd \cite{cd106}; 
vi) $\beta\beta$ and $\beta$ decay modes in $^{48}$Ca \cite{ca48}; 
vii) 2EC2$\nu$ in $^{ 136}$Ce and in $^{138}$Ce and $\alpha$ decay in $^{142}$Ce \cite{Ce}; 
viii) 2$\beta^+0\nu$ and EC$\beta^+0\nu$ decay modes in $^{130}$Ba \cite{Ba}.
Moreover, the data taken with a low background CdWO$_4$ scintillator are under analysis and 
a data taking with a ZnWO$_4$ detector is in progress. 
Presently, a molybdenum sample of 1 kg mass, enriched in the $^{100}$Mo isotope at 99.5\%, is in measurement
in the low-background set-up with 4 HPGe detectors of the LNGS facility; this experiment \cite{armonia}
will test previous reports on observation of $2\beta2\nu$ decay of $^{100}$Mo to the first excited 
$0^+_1$ level of $^{100}$Ru.
New data taking are foreseen also e.g. with a
new larger BaF$_2$ scintillator and with Cadmium enriched in $^{106}$Cd. 

In particular, by using the DAMA/LXe set-up filled with Kr-free Xenon
gas containing 17.1\% of $^{134}$Xe and 68.8\%  of $^{136}$Xe
double beta decay modes in $^{134}$Xe and $^{136}$Xe have been investigated \cite{LXe-bb1,LXe-bb2};
the measurements have been carried out over 8823.54 h.
A joint analysis of the $0\nu\beta\beta$ decay
mode in $^{134}$Xe and in $^{136}$Xe (as suggested in ref. \cite{Faes})
has also been carried out. The obtained new lower limits 
for the $0\nu \beta\beta (0^+ \rightarrow 0^+)$
decay mode in $^{134}$Xe and in $^{136}$Xe (90\% C.L.) are
$T_{1/2} = 5.8 \cdot 10^{22}$ y and  $T_{1/2} = 1.2 \cdot 10^{24}$ y, respectively.
They correspond to a limit value on {\it effective light Majorana
neutrino mass} ranging from 1.1 eV to 2.9 eV (90\% C.L.), depending on the
adopted theoretical model. For the neutrinoless double beta
decay with Majoron ($M$) in the $^{136}$Xe isotope
the obtained limit is $T_{1/2} = 5.0 \cdot 10^{23}$ y (90\% C.L.) and 
for the $2\nu \beta\beta (0^+\rightarrow 0^+)$ and
the $2\nu \beta\beta (0^+ \rightarrow 2^+)$ decay
modes in $^{136}$Xe the limits at 90\% C.L. are
$1.0 \cdot 10^{22}$ y and $ 9.4 \cdot 10^{21}$ y, respectively.
The experimental limit on the $2\nu \beta\beta (0^+\rightarrow 0^+)$
decay mode is in the range of the theoretical estimate by
\cite{Stau} ($2.11 \cdot 10^{22}$ y) and about a factor 5 higher than
that of ref. \cite{Cau}; on the other hand, similar
theoretical estimates suffer of the large uncertainties typically associated to the
calculations of the nuclear matrix elements and the overcoming of these theoretical estimates
has also been confirmed by the Baksan Xenon experiment \cite{XeBak}. 

Among the other results we want to point out here just our efforts in preliminary studies 
on $\beta^+\beta^+$/$EC\beta^+$/$ECEC$ decay 
modes since potentially they can offer both a large number of relatively poorly 
explored double beta decaying isotopes and the 
possibility of a good signature for the processes. In particular, 
neutrinoless double positron (0$\nu2\beta^+$) decay, $\beta^+$/EC decay and EC/EC decay can give, in principle,
the same information as 0$\nu2\beta^-$ decay. In addition,
they can play a key role in recognising the main
production mechanism, strongly depending their calculated
half-lives on whether the decay is dominated
by Majorana neutrino mass or by right-handed admixtures
in weak interactions [5]. On the other hand,
even the non-observation of the 0$\nu2\beta^+$ decay would
provide very useful additional and complementary
information.
The current level of experimental sensitivity is modest
compared with those already achieved in the $2\beta^-$ decay 
searches, although the present theoretical expectations
for the half-lives of some $2\beta^+$ candidate isotopes
are favourable. To reach a suitable sensitivity the
use of enriched samples is needed to effectively exploit
the potentiality of the $2\beta^+$ processes, which have
the advantage of a very clear signature. In fact, e.g., in
the 0$\nu\beta^{+}/EC$ decay the monoenergetic positron will
give two 511 keV annihilation $\gamma$-rays in addition to
characteristic X-rays.

A first step in this direction was done in ref. \cite{cd106} where we present
the results of a study on the $2\beta^+$ decay processes in
$^{106}$Cd, which is a one of the best candidate nuclides
because of its rather high Q value ($2\beta^+$), (2771 $\pm$ 8) keV
and to the favourable theoretical estimates
for half-lives. In this experiment, samples of metallic cadmium
enriched in $^{106}$Cd at $\simeq$ 68\% (total mass $\simeq$ 154 g)
were placed between two low
background NaI(Tl) crystals in the DAMA/R\&D set-up;
measurements were performed during 4321 h.
New improved limits on the halflives
of the various decay modes have been obtained \cite{cd106}; they were in the range
(0.3 -- 4) 10$^{20}$ y (90\% C.L.), that is, they were significantly
higher (by factor 6 to 60) than those previously
published for the same nuclide, approaching the theoretical
estimates. In particular, the calculated T$_{1/2}$ values for the $2\nu\beta^+/EC$ decay of $^{106}$Cd 
are within $(8 - 40) 10^{20}$ y,
which are not so far from our limit for this process.
These results have strongly supported the interest in
pursuing a further enhancement
of the experimental sensitivity by
additional purification of the enriched
cadmium samples and the growth from them
of a $^{106}$CdWO$_4$ scintillator, which could be used as
source-detector and, in combination with low background
NaI(Tl) crystals, could allow to identify signals. At present some work is in progress in this direction.

Among the other efforts, we remind the results on the neutrinoless double electron 
capture of $^{40}$Ca by using low background CaF$_{2}$(Eu) 
scintillators in the R\&D set-up. The obtained new and highly improved T$_{1/2}$ limit: 3.0 $\times$ 10$^{21}$ y,
has demonstrated the feasibility and perspectives of developed CaF$_{2}$(Eu) scintillators for high sensitive investigations of the 
double beta decay modes. 

On the same line, BaF$_{2}$ \cite{Ba} and CeF$_{3}$ \cite{Ce} scintillators have been used in preliminary investigations both 
on the level of their residual radiopurity and on the possibility of investigation of some double beta decay modes in $^{130}$Ba 
and of double electron capture processes in $^{136}$Ce and in $^{138}$Ce.

The potentiality of the coincidence technique to search for double beta  
decay processes in $^{130}$Ba has been studied by using a 
BaF$_{2}$ detector and two low background NaI(Tl) in the DAMA/R\&D set-up.
The performances of the detector have preliminarily been investigated. 
Then, as a result, new limits (90\% CL) on the lifetimes of 
the following double beta decay processes in  $^{130}$Ba have been determined: 
$\tau^{2\beta^+ 0\nu (0^+ \rightarrow 0^+)}$ $>$ $ 1.1 \times 10^{17}$ y,
$\tau^{2\beta^+ 0\nu (0^+\rightarrow 2^+_1)}$ $>$ $ 1.5\times 10^{17}$ y,
$\tau^{EC\beta^+ 0\nu (0^+\rightarrow 0^+)}$ $>$ $ 2.0 \times 10^{17}$ y,
$\tau^{EC\beta^+ 0\nu (0^+ \rightarrow 2^+_1)}$ $>$ $ 1.6 \times 10^{17}$ y,
$\tau^{EC\beta^+ 0\nu (0^+ \rightarrow 2^+_2)}$ $>$ $ 2.0 \times 10^{17}$ y and
$\tau^{EC\beta^+ 0\nu (0^+\rightarrow 4^+)}$ $>$ $ 1.1 \times 10^{17}$ y.
The new experimental limits obtained for the $0\nu (0^+ 
\rightarrow 0^+)$ decay modes represent a signiﬁcant improvement with respect to those
previously available for this isotope, while those involving excited levels have been set 
for the first time. Measurements with 
a new BaF$_{2}$ detector already at hand are foreseen.

As regards CeF$_{3}$, because of the still relatively important residual 
radioactive contamination, the obtained limits \cite{Ce} are: 
T$^{2\nu2K}_{1/2}$($^{136}$Ce) $\ge$ 
2.7(4.5) $\times$ 10$^{16}$ y and T$^{2\nu2K}_{1/2}$($^{138}$Ce) $\ge$ 
3.7(6.1) $\times$ 10$^{16}$ y
respectively, at 90\%(68\%) C.L. This work has shown that signiﬁcant improvements
in the sensitivity to rare processes in Ce
isotopes can be obtained by exploiting the investigated
techniques, but also that it is a crucial point in the achievement of 
more competitive results 
to overcome the limitation arising from the residual
radioactive impurities in this kind of detector.
Puriﬁcation of scintillators from actinides and
their daughters to the level of few $\mu$Bq/kg could
decrease the background at least to few
counts/(day keV kg). This together with the use
of isotopical enrichment in the production of CeF$_{3}$ scintillators
would improve the experimental sensitivity
at the level expected from the theoretical
predictions.

\vspace{-0.5cm}
\section{Conclusions}

Significant results have been obtained by the DAMA collaboration in the investigation
of several rare processes and large efforts are in progress to further improve 
this understanding. New results will be available in near future.

\vspace{-0.5cm}
\section*{Acknowledgement}

I thank the organizer of this workshop for offering me this opportunity of discussion
and all the colleagues working in the framework of the DAMA activity.

\end{document}